\documentclass[aps,prl,reprint,balancelastpage,nofootinbib,preprintnumbers,superscriptaddress]{revtex4-1}

\usepackage{slashed}
\usepackage{bm}
\usepackage{latexsym,amssymb,amsmath,float,url,mathrsfs}
\usepackage{latexsym}
\usepackage{graphicx}
\usepackage{epstopdf}
\usepackage{amsmath}
\usepackage{comment}
\usepackage{subfigure}
\usepackage{natbib}
\usepackage{hyperref}
\usepackage{pifont}
\usepackage{blindtext}
\usepackage{adjustbox}
\usepackage{multirow}
\usepackage{tabularx}
\usepackage[table]{xcolor}
\usepackage{orcidlink}
\usepackage{tikz}
\usetikzlibrary{tikzmark}
\usepackage{fontawesome}

\usepackage{blkarray}
\usepackage{graphicx}
\usepackage{amsfonts}
\usepackage{soul}
\usepackage{amssymb}
\usepackage{amsmath}
\usepackage{cancel}
\usepackage{tcolorbox}

\usepackage{graphics,appendix,afterpage,makecell} 

\definecolor{oucrimsonred}{rgb}{0.6, 0.0, 0.0}
\definecolor{persianblue}{rgb}{0.11, 0.22, 0.73}
\definecolor{forestgreen}{rgb}{0.13,0.35,0.13}
\definecolor{lightgray}{rgb}{0.83, 0.83, 0.83}
 \hypersetup{colorlinks, citecolor=oucrimsonred, linkcolor=black, urlcolor=oucrimsonred}
\definecolor{cornellred}{rgb}{0.7, 0.11, 0.11}
\definecolor{navyblue}{rgb}{0.0, 0.0, 0.5}
\definecolor{amethyst}{rgb}{0.6, 0.4, 0.8}
\definecolor{yellow}{rgb}{1.0, 1.0, 0.0}
\definecolor{firebrick}{rgb}{0.7, 0.13, 0.13}
\definecolor{tangerineyellow}{rgb}{1.0, 0.8, 0.0}
\definecolor{deepfuchsia}{rgb}{0.76, 0.33, 0.76}
\definecolor{amber}{rgb}{1.0, 0.75, 0.0}
\definecolor{VioletRed4}{rgb}{0.55, 0.13, .32}
\definecolor{indiagreen}{rgb}{0.07, 0.53, 0.03}
\definecolor{VioletRed4}{rgb}{0.55, 0.13, .32}
\newcommand{\be}{\begin{equation}}
\newcommand{\ee}{\end{equation}}
\newcommand{\bea}{\begin{equation} \begin{aligned}}
\newcommand{\eea}{\end{aligned} \end{equation}}

\definecolor{oucrimsonred}{rgb}{0.6, 0.0, 0.0}
\newcommand\vertarrowbox[3][6ex]{%
  \begin{array}[t]{@{}c@{}} #2 \\
  \left\uparrow\vcenter{\hrule height #1}\right.\kern-\nulldelimiterspace\\
  \makebox[0pt]{\scriptsize#3}
  \end{array}%
}

\definecolor{verdechiaro}{rgb}{0.6,1,0.6}
\definecolor{giallochiaro}{rgb}{1,1,0.6}
\definecolor{bluscuro}{rgb}{0.15, 0.2, 0.9}
\definecolor{verdes}{rgb}{0.1, 0.5, 0.1}%
\definecolor{tangerineyellow}{rgb}{1.0, 0.8, 0.0}

\definecolor{americanrose}{rgb}{1.0, 0.01, 0.24}
\definecolor{cobalt}{rgb}{0.0, 0.28, 0.67}
\definecolor{brandeisblue}{rgb}{0.0, 0.44, 1.0}
\definecolor{mycolor}{rgb}{0.0, 0.0, 0.5}
\definecolor{oxfordblue}{rgb}{0.0, 0.13, 0.28}
\definecolor{azure}{rgb}{0.0, 0.5, 1.0}
\definecolor{turquoiseblue}{rgb}{0.0, 1.0, 0.94}
\newtcolorbox{mynewbox}[1]{colback=white!5!white,colframe=azure!75!black,fonttitle=\bfseries,title=#1}
\newtcolorbox{mybox}{colback=mycolor!5!white,colframe=azure!75!black}
\newtcolorbox{mynamedbox}[1]{colback=mycolor!5!white,colframe=azure!75!black,title=#1}
\definecolor{venetianred}{rgb}{0.78, 0.03, 0.08}
\newtcolorbox{mynamedbox1}[1]{colback=venetianred!5!white,colframe=venetianred!80!black,title=#1}
\newtcolorbox{mynamedbox2}[1]{colback=azure!5!white,colframe=azure!80!black,title=#1}

\definecolor{verdes}{rgb}{0.1, 0.5, 0.1}%
\definecolor{cornellred}{rgb}{0.7, 0.11, 0.11}

\definecolor{VioletRed4}{rgb}{0.55, 0.13, .32}

\hypersetup{
     colorlinks   = true,
     citecolor    = violet,
     urlcolor     = violet,
     linkcolor    = violet}

\definecolor{rossocorsa}{rgb}{0.83, 0.0, 0.0}

\usepackage[normalem]{ulem}

\newcommand{\papertitle}{Nonlinear Tails of Gravitational Waves in Schwarzschild Black Hole
Ringdown  }

\begin{document}

\title[]{\papertitle}

\author{Alex Kehagias\orcidlink{}}
\affiliation{Physics Division, National Technical University of Athens, Athens, 15780, Greece}
\affiliation{CERN, Theoretical Physics Department, Geneva, Switzerland}

\author{Antonio Riotto\orcidlink{0000-0001-6948-0856}}
\affiliation{Department of Theoretical Physics and Gravitational Wave Science Center,  \\
24 quai E. Ansermet, CH-1211 Geneva 4, Switzerland}


\begin{abstract}
\noindent
 Schwarzschild black holes evolve toward their static configuration by emitting gravitational waves, which decay over time following a power law at fixed spatial positions. We derive this power law analytically for the second-order even gravitational perturbations, demonstrating that it is determined by the fact that the second-order source decays as the inverse square of the distance. Quadratic gravitational modes with multipole $\ell$ decay according to a law $\sim t^{-2\ell - 1}$, in contrast to the linear Price law scaling $\sim t^{-2\ell - 3}$. Consequently, nonlinear tails may persist longer than their linear counterparts.

\end{abstract}

\maketitle

\noindent\textbf{Introduction --} 
The advent of gravitational wave (GW) astronomy has ushered in a new era for conducting precise tests of strong-field gravity \cite{Berti:2015itd,Berti:2018vdi,Franciolini:2018uyq,LIGOScientific:2020tif}. Ground-based detectors, such as LIGO, VIRGO, and KAGRA, alongside the upcoming space-based mission LISA, are achieving sensitivities that enable detailed studies of black hole (BH) merger ringdowns \cite{Berti:2005ys,KAGRA:2013rdx,LIGOScientific:2016aoc,KAGRA:2021vkt,LIGOScientific:2023lpe}. A critical aspect of this exploration involves understanding the late-time behavior of perturbations around BHs, where both linear and nonlinear effects leave distinctive imprints on the emitted GWs.

For years, linear BH perturbation theory has provided the foundation for modeling BH ringdowns, yielding the well-established result that, at late times, BHs relax through a sequence of exponentially damped oscillations, known as quasinormal modes (QNMs), followed by an inverse power-law decay with time. This is famously represented by Price's scaling: massless fields in a non-spinning BH background decay at fixed spatial locations as $\sim t^{-2\ell-3}$ at very late times, where $\ell$ is the corresponding multipole number \cite{Price:1971fb,Price:1972pw,Leaver:1986gd,Gundlach:1993tp,Gundlach:1993tn,Ching:1995tj,Chandrasekhar:1975zza,Martel:2005ir,Barack:1998bw,Berti:2009kk}.

To accurately model the ringdown waveforms of BHs, significant efforts have been dedicated to refining our understanding of the QNM spectra and Price tails \cite{Ching:1994bd,Krivan:1996da,Krivan:1997hc,Krivan:1999wh,Burko:2002bt,Burko:2007ju,Hod:2009my,Burko:2010zj,Racz:2011qu,Zenginoglu:2012us,Burko:2013bra,Baibhav:2023clw}. Linear predictions have been rigorously validated by numerical relativity simulations and increasingly precise GW observations. Recently, there has been a surge of interest in the nonlinear evolution of BH perturbations \cite{Gleiser:1995gx,Gleiser:1998rw,Campanelli:1998jv,Garat:1999vr,Zlochower:2003yh,Brizuela:2006ne,Brizuela:2007zza,Nakano:2007cj,Brizuela:2009qd,Ripley:2020xby,Loutrel:2020wbw,Pazos:2010xf,Sberna:2021eui,Redondo-Yuste:2023seq,Mitman:2022qdl,Cheung:2022rbm,Ioka:2007ak, Kehagias:2023ctr,Khera:2023oyf,Bucciotti:2023ets,Spiers:2023cip,Ma:2024qcv,Zhu:2024rej,Redondo-Yuste:2023ipg,Bourg:2024jme,Lagos:2024ekd,Perrone:2023jzq,Kehagias:2024sgh,Bucciotti:2025rxa,bourg2025quadraticquasinormalmodesnull,BenAchour:2024skv,Kehagias:2025ntm}, uncovering phenomena absent in linear treatments, such as quadratic QNMs and nonlinear power-law tails.

These nonlinear tails are of particular interest, as they naturally emerge from the outgoing QNM profiles present in the ringdown dynamics \cite{Okuzumi:2008ej,Lagos:2022otp}, with their origins closely tied to the nonlinear nature of gravity. Recent numerical advancements have identified the power-law behaviors associated with these nonlinear tails for the Weyl scalar, which are distinct from Price tails \cite{Cardoso:2024jme}. For instance, it has been shown that the power-law of the $\ell=4$ mode generated at second order from two $\ell=2$ modes decays at large times as $\sim t^{-10}$, suggesting a more general law of the form $\sim t^{-2\ell-2}$. Moreover, recent 3+1-dimensional numerical relativity simulations of BH mergers have provided evidence for power-law tails that diverge from the Price tail \cite{DeAmicis:2024eoy,Ma:2024hzq}. These developments provide insight into the rich phenomenology of nonlinear effects in BH perturbation theory and highlight the necessity for new experimental efforts, such as GW waveform modeling, to detect their corresponding signatures.

The aim of this letter is to offer a simple  analytical understanding of the nonlinear tails of GWs emitted during the Schwarzschild BH ringdown. As we will demonstrate, the power law is fully governed by the fact that the second-order source decays as $\sim 1/r^2$ at large distances from the BH  and, as a consequence,   the corresponding Green function must be  solved in (a slightly) curved spacetime.
\vskip 0.5cm
\vspace{-5pt}\noindent\textbf{Spectroscopy for a Schwarzschild BH --} 
Our starting point is a Schwarzschild BH with mass $M$ whose spacetime is described by the   metric (setting $G_N=1$)

\be
\label{metric}
{\rm d} s^2=-f(r){\rm d} t^2+\frac{{\rm d} r^2}{f(r)}+r^2{\rm d}\Omega_2^2, \,\,\,f(r)=1-2M/r,
\ee
Adopting the Regge-Wheeler-Zerilli formalism \cite{Regge:1957td,Zerilli:1970wzz} for the first-order metric perturbations in the Schwarzschild
spacetime and separating angular variables with tensor harmonics
of indices $(\ell,m)$,  the equations are divided into the even and odd parity parts.  In this paper we limit ourselves to the the even parity mode, corresponding to the  Zerilli equation at first-order. Even though there are seven equations for the even parity part of the metric perturbation, at first-order one can reduce the dynamics to a single Zerilli equation for a function $\psi^{(1)}_{\ell m}(r,t)$ in the so-called Regge-Wheeler (RW) gauge

\begin{eqnarray}
&&\left[-\frac{\partial^2}{\partial t^2}+\frac{\partial^2}{\partial r_*^2}-V(r)\right]\psi_{\ell m}^{(1)}(r,t)=0,\nonumber\\
V(r)&=&f(r)
\frac{2\lambda^2(\lambda+1)r^3+6\lambda^2 M r^2+18\lambda M^2 r+18 M^3}{r^3(\lambda r+3 M)^2},\nonumber\\
&&
\end{eqnarray}
where 

\begin{eqnarray}
r_*&=&r+2M\ln(r/2M-1),\nonumber\\
\lambda&=&\frac{1}{2}(\ell-1)(\ell+2).
\end{eqnarray}
The  evolution is determined by a  retarded Green’s function solving the equation

\be
\left[-\frac{\partial^2}{\partial t^2}+\frac{\partial^2}{\partial r_*^2}-V(r)\right]G(t-t';r,r')=\delta(t-t')\delta(r-r'),
\ee
under the boundary condition  $G(t-t';r,r')=0$ for $t<t'$. The retarded Green function receives three contributions, $G=(G_F+G_Q+G_B)$. 
At linear order and 
for $t'\geq t- |r- r'|$, the Green’s
function $G\simeq  G_F$ behaves as in  flat background: a wave 
arrives at $(t,r)$ without passing through the potential barrier, and thus without being reflected or deformed. For $t'\leq t - |r- r'|$,  the Green’s function $G \simeq G_Q$ oscillates with QNM frequencies: a wave generated in this region
pass through the potential and is  deformed into
a first-order QNM. Finally, the branch-cut part $G_B$ describes the
backscattering of waves by the long-range part of the potential barrier and is responsible for the Price's tail. At the linear order  the contribution of $G_B$  is usually small
compared to the other contributions \cite{Ching:1995tj}, and
 one may  neglect it.
In the Supplementary Material we offer a self-similar and quantum mechanical derivation of the linear Price's tail which is alternative  to the usual Green function branch-cut argument.

 At second-order,   perturbation theory becomes much more involved, but there are general considerations which we may propose to arrive at some universal results. Consider for instance the multipole $\ell=4$ which can be constructed out of the linear modes $(2,\pm 2)$. Following closely Ref. \cite{Nakano:2007cj}, one can write the second-order equation for $\psi_{4\pm 4 }^{(2)}(t,r)$ as

 \begin{equation}
    \left[-\frac{\partial^2}{\partial t^2}+\frac{\partial^2}{\partial r_*^2}-V(r)\right]\psi^{(2)}_{4\pm 4 }(r,t)= S_{4\pm 4 }(r,t),
 \end{equation}
where $
S_{4\pm 4 }(r,t)$ is the regularized quadratic source built up with  the linear modes $(2,\pm 2)$. The key point is that it  behaves as $\sim 1/r^2$ at large radii (and as $(r-2M)$ close to the horizon) 

\be
S_{4\pm 4 }(r\gg 2M,t)\sim \frac{1}{r^2}\left[\psi^{(2)}_{2\pm 2 }(r,t)\right]^2.
\ee
The behaviour   is in fact valid
for any multipole, and in the following it will suffice to use the following   expression 

\be
\label{a}
S_{\ell\pm \ell }(r\gg 2M,t)\sim \frac{e^{-2i\omega_\ell(t-r_*)}}{r^2}, 
\ee
where $\omega_\ell$ is the QNM of the multipole ($\ell,\pm\ell)$. It represents the starting point for our considerations on the nonlinear tails.

\vskip 0.5cm
\vspace{-5pt}\noindent\textbf{Nonlinear tails --} 
 As we have already mentioned, at  first-order in perturbation theory, the QNMs arise from the QNM component \( G_Q \) of the Green’s function, while the Price's power-law tails come from the branch-cut component \( G_B \). This indicates that both are generated in a region in which the potential $V(r)$ is important: the first-order QNMs are excited near the peak of the potential barrier, and the first-order tails emerge from the long-range decay of the potential or, equiva-
lently, the fact that perturbations in a non-flat space-time travel not
only on the light cone, but also inside it.
 
 In contrast, the second-order  power-law tail originates from the flat part \( G_F \) of the Green’s function, meaning they are produced in asymptotically ``flat regions", and from the 
 edge of the source distribution at $(t'-r_*')\simeq u_0$ with some finite support.
 It might be surprising that a non-oscillating tail can be produced by oscillating QNMs, but it can be understood as follows \cite{Okuzumi:2008ej}:  an observer at position \( (t, r_*) \) sees the edge $(t'+r_*')$ extending from \(  ( t - r_*) \) to \( ( t + r_*) \) and  cannot  detect the edge for \( (t'+r_*')  < t - r_* \), as waves generated in this region are partially scattered by the potential barrier before reaching the observer. The first-order perturbations $\psi^{(1)}_{\ell m}(t',r_*')\sim e^{-i\omega(t'-r_*')}$ are constant along the edge $(t'-r_*')\simeq u_0$ and, therefore, they do not exhibit oscillations, thus leading to  long-lasting  tails \cite{Okuzumi:2008ej}.

Now we arrive at a  critical observation. Since the source at large values of $r$ decays at $\sim 1/ r^2$, and since

\be
V(r\gg 2M)\simeq  \frac{\ell(\ell+1)}{r^2},
\ee
it is clear that the flat Green function $G_F$ may not be evaluated neglecting such a leading term as it scales with the same power of the source. In other terms, the 
flat part $G_F$ of the Green function must be calculated taking the centrifugal term  in the potential. 

Imagine that  QNMs
turn on at a given time $t_0$ at a given point $r_0$ (typically the photon ring $r_0\simeq 3M$ for $\ell\gg 1$) and 
 moves outward at light speed, with a finite support in term of the retarded time $ u'\equiv (t'-r'_*)\simeq u_0$. On the other hand   the advanced time $v'\equiv (t'+r_*')$ ranges from $(t-r_*)$ to $(t+r_*)$. We further  assume that the  observer is located at points   farther apart in the time direction than in the radial 
direction, that is $t\gg r_*$. Under these circumstances the source
acquires the form

\be
S_{\ell\pm \ell }(u,v)\sim \frac{e^{-2i\omega_\ell u_0}}{r^2(u_0,v)}, 
\ee
up to a multiplicative constant.
The nonlinear tail
is therefore dictated by the integral 

\be
\label{fund}
{\cal I}(t,r)=\int_{t-r_*(r)}^{t+r_*(r)}\frac{{\rm d}v'}{r^2(u_0,v')}\, G_F(t,r;u_0,v'),
\ee
where one has also to   account for the fact that at large radii $r_*$ is not exactly equal to $r$ and subleading terms are relevant to catch the nonlinear tails.
The closed form of $G_F$ is obtained in the Supplementary Material. For instance, for $\ell=2$, it reads

\begin{eqnarray}
G_F(t,r;t',r')&=&\sqrt{\frac{\pi}{2}}\frac{1}{16 r^2 {r'}^2}\left[(t-t')^4 + 3 r^4+2 r^2 {r'}^2\right.\nonumber\\
&+&\left.3 {r'}^4-6(t-t')^2(r^2+{r'}^2)\right], \nonumber
\end{eqnarray}
 leading to

\be
{\cal I}(t\gg r)\simeq \frac{32}{15}\sqrt{\frac{\pi}{2}}\frac{r^3}{t^6}.
\ee
For $\ell=4$, using the Green function given in the Supplementary Material, we find

\be
{\cal I}(t\gg r)\simeq \frac{512}{315}\sqrt{\frac{\pi}{2}}\frac{r^5}{t^{10}},
\ee
which reproduces the scaling found numerically in Ref. \cite{Cardoso:2024jme}.
Furthermore, by using the general form of the Green function in Eq. (\ref{GG1}), our general finding is 

\be
{\cal I}(t\gg r)\simeq \frac{(-1)^\ell 2^{2\ell+1} \ell! (\ell-1)!}{(2\ell+1) (2\ell-1)!}\sqrt{\frac{\pi}{2}}\, \frac{r^{\ell+1}}{t^{2\ell+2}}. 
\label{Igen}
\ee
Let us notice here that a similar result holds for the Reissner-N\"ordstrom BH with metric of the form (\ref{metric}) with $f=1-2M/r+Q^2/r^2$. In this case, using the corresponding relation $r_*(r)$ in the integral (\ref{fund}), we get 

\be
{\cal I}_{\rm RN}(t\gg r)\simeq C_\ell\, \frac{(-1)^\ell 2^{2\ell+1} \ell! (\ell-1)!}{(2\ell+1) (2\ell-1)!}\sqrt{\frac{\pi}{2}}\,  \frac{r^{\ell+1}}{t^{2\ell+2}}, 
\label{Igen}
\ee
where 
\be
C_\ell=\sum_{k=0}^{\left[\frac{\ell}{2}\right]}\frac{(l-k)!}{(\ell-2k)! k!} (2M)^{\ell-2k} Q^{4k}.
\ee
 There is one more passage to perform though to arrive at the final result.

\vskip 0.5cm
\vspace{5pt}\noindent\textbf{From the RW  to the asymptotic flat gauge   --} 
The last necessary step we have to take in order to 
 obtain the correct power-law tail of the gravitational waveform is to go
to an asymptotic flat  (AF) gauge from the RW gauge. This is because the Regge-Wheeler-Zerilli formalism that we
have employed is under the RW gauge and this gauge is not asymptotically flat.
 Following Ref. \cite{Nakano:2007cj}, we can write the fundamental modes  at infinity in the asymptotic flat  gauge  at first- and second-order for the modes $(2,\pm 2)$ and $(4,\pm 4)$ as

\begin{eqnarray}
h^\text{(1) \hspace{-0.1cm}{\rm AF} }_{2,\pm 2}&\simeq &\frac{1}{r}\psi^{(1)}_{2,\pm 2},\nonumber\\
\frac{\partial h^\text{(2) \hspace{-0.2cm} AF}_{4,\pm 4}}{\partial t}&\simeq &\frac{1}{r}\psi^{(2)}_{4,\pm 4}+\frac{\sqrt{70}}{1512\sqrt{\pi}r}\left[27 \psi^{(1)}_{2,\pm 2}\frac{\partial^2 \psi^{(1)}_{2,\pm 2}}{\partial t^2}\right.\nonumber\\
&+&\left.24\left(\frac{\partial \psi^{(1)}_{2,\pm 2}}{\partial t}\right)^2+4 M\left(\frac{\partial \psi^{(1)}_{2,\pm 2}}{\partial t}\right)\frac{\partial^2 \psi^{(1)}_{2,\pm 2}}{\partial t^2}\right].\nonumber\\
&&
\end{eqnarray}
From this expression, we deduce that the leading power-law tail for the gravitational waveform for the even modes is

\be
h^\text{(2) \hspace{-0.1cm}{\rm AF} }_{\ell,\pm m}
\sim \int^t{\rm d} t'\,\frac{1}{r}\psi^{(2)}_{\ell m}\sim 
t^{-2\ell-1}.
\ee
This is our final result for the  nonlinear tail
of the even gravitational  degrees of freedom.

\vskip 0.5cm
\vspace{5pt}\noindent\textbf{Conclusions  --} 
It is well-established that, when a BH reaches its static configuration, linearized fluctuations at fixed spatial positions decay over time according to the Price's power law. In this Letter, we have provided an analytical explanation for the power-law tail that governs the decay of nonlinear perturbations at second order in perturbation theory. The exponent of this power law is determined by the fact that the second-order source decays as the inverse of the square of the distance, which may result in a significantly longer decay time compared to the linear case. Specifically, the quadratic even modes decay as $\sim t^{-2\ell-1}$, in contrast to the linear Price's scaling  $\sim t^{-2\ell-3}$. In other words, the linear perturbation theory might be  inadequate for predicting the late-time evolution of the ringdown, necessitating the inclusion of higher-order corrections.
\vskip 0.3cm
\centerline{\it Note Added}
\vskip 0.2cm
\noindent
When finalizing this work, we became aware of Ref. \cite{Ling:2025wfv}, which addresses the nonlinear tail of an interacting scalar field in a Schwarzschild spacetime. In contrast to this reference, we have focused on the nonlinear tail of the second-order gravitational degrees of freedom. Nevertheless, where overlap occurs, our results are fully consistent.
\vskip 0.1cm
\begin{acknowledgments}
\vspace{5pt}\noindent\emph{Acknowledgments --}
  A.R.  acknowledges support from the  Swiss National Science Foundation (project number CRSII5\_213497) and from  the Boninchi Foundation through  the project ``PBHs in the Era of GW Astronomy''.
\end{acknowledgments}

\bibliography{main}
\clearpage
\maketitle
\onecolumngrid
\begin{center}
\textbf{\large \papertitle} 
\\ 
\vspace{0.05in}
{Alex Kehagias  and Antonio Riotto}
\\ 
\vspace{0.2in}
{ \large \it Supplementary Material}
\end{center}
\onecolumngrid
\setcounter{equation}{0}
\setcounter{figure}{0}
\setcounter{section}{0}
\setcounter{table}{0}
\setcounter{page}{1}
\makeatletter
\renewcommand{\theequation}{S\arabic{equation}}
\renewcommand{\thefigure}{S\arabic{figure}}
\renewcommand{\thetable}{S\arabic{table}}

\vspace{-0.5cm}
\section{Linear tails}
\noindent
\label{supp:1}
In this Supplementary Material we offer an alternative derivation for the Price's tail for the odd modes. The corresponding metric perturbations for  the Schwarzschild BH are the functions $h_{\ell m}^{(1)}(t,r)$ and $h_{\ell m}^{(0)}(t,r)$ and they  are determined by the solution $Q_{\ell m}(t,r)$  of the RW equation as 
\begin{eqnarray}
h_{\ell m}^{(1)}=\left(1-\frac{2M}{r}\right)^{-1} r\, Q_{\ell m}, \qquad \frac{\partial h_{\ell m}^{(0)}}{\partial t}=\frac{\partial(r\, Q_{\ell m})}{\partial r_*}, 
\end{eqnarray}
where $Q_{\ell m}(t,r)$ is the time derivative of $q_{\ell m}(r,t)$, such that \cite{Price:1971fb}
\begin{eqnarray}
q_{\ell m}(r,t)=\int_\infty^t \,{\rm d}t'\,Q_{\ell m}(r,t').
\end{eqnarray}
At large radii, $q_{\ell m}$ satisfies the equation
\begin{eqnarray}
\frac{\partial^2 q_{\ell m}}{\partial t^2}-\frac{\partial^2 q_{\ell m}}{\partial r^2}+\frac{\ell(\ell+1)}{r^2}q_{\ell m}=0. \label{qq}
\end{eqnarray} 
This differential equation is invariant under the rescaling $(t,r)\to (\lambda t,\lambda r)$ with a constant $\lambda$,  so that we we may look for self-similar  solutions of the form 
\begin{eqnarray}
 q_{\ell m}(t,r)=\frac{1}{t^a} \phi_{\ell m}(z), \qquad z=\frac{t}{r}.
 \end{eqnarray} 
It then turns out that $q_{\ell m}(z)$ satisfies 
\begin{eqnarray}
(1-z^2) \phi_{\ell m}''-\frac{2}{z} (a+z^2)\phi'_{\ell m}+\left(\ell(\ell+1)+\frac{a(a+1)}{z^2}\right)\phi_{\ell m}=0.
\end{eqnarray}
Expressing $\phi_{\ell m}$ as 
\begin{eqnarray}
\phi_{\ell m}(z)= \frac{z^a}{(z^2-1)^{a/2}} f_{\ell m}, 
\end{eqnarray}
we find that $f_{\ell m} $ satisfies 
\begin{eqnarray}
(1-z^2)f_{\ell m}''-2zf_{\ell m}'+\left(\ell(\ell+1)-\frac{a^2}{1-z^2}\right)f_{\ell m}=0.  \label{flm}
\end{eqnarray}
Eq. (\ref{flm}) is the associated Legendre differential equation and it is solved by the associated Legendre functions $P_\ell^{a}(z)$ and $Q_\ell^a(z)$ as 
\begin{eqnarray}
f_{\ell m}=A_{\ell m}  P_\ell^{a}(z)+B_{\ell m} Q_\ell^{a}(z). 
\label{flm1}
\end{eqnarray}
As we are interested in the region $z>1$, we may use the following expressions of $P_\ell^{a}(z)$ and $Q_\ell^a(z)$
\begin{eqnarray}
P_\ell^{a}(z)&=& \frac{2^{-\ell -1}\pi^{-\frac{1}{2}}\Gamma(-\frac{1}{2}-\ell)z^{-\ell+a-1}}{(z^2-1)^{a/2}\Gamma(-\ell-a)}\, {}_2F_1\left(\frac{1+\ell-a}{2},1+\frac{\ell-a}{2},\ell+\frac{3}{2}, \frac{1}{z^2}\right)\nonumber \\
&&\hspace{1.5cm} +\frac{2^\ell\, \Gamma(\frac{1}{2}+\ell)z^{\ell+a}}{(z^2-1)^{a/2}\Gamma(1+\ell -a)} \, {}_2F_1\left(-\frac{\ell+a}{2},\frac{1-\ell-a}{2},\frac{1}{2}-\ell, \frac{1}{z^2}\right),\label{P}
\\
Q_\ell^a(z)&=&e^{ia\pi} 2^{-\ell-1}\pi^{\frac{1}{2}}
\frac{\Gamma(\ell+a+1)}{\Gamma(\ell+\frac{3}{2})}   z^{-\ell-a-1}(z^2-1)^{\frac{a}{2}}\, {}_2F_1\left(1+\frac{\ell+a}{2},\frac{1+\ell+a}{2},\ell+\frac{3}{2}, \frac{1}{z^2}\right). \label{Q}
\end{eqnarray}
On the other hand, the static problem of Eq. (\ref{qq}), i.e. $q_{\ell m}=q_{\ell m}(r)$, is solved by the two independent modes
\begin{eqnarray}
q^{(1)}_{\ell m}(r)=r^{-\ell}, \qquad q^{(2)}_{\ell m}(r)=r^{\ell+1}.
\end{eqnarray}
The dominant one for large $r$ is the $r^{\ell+1}$ solution which is obtained for large $z$ (large $t$ and fixed $r$) if the second term in the right-hand side of Eq.(\ref{P}) vanishes. This is possible if $\Gamma(1+\ell-a)$ has a pole, which is possible when 
\begin{eqnarray}
 a=\ell+1+n, \qquad n=0,1,\cdots.
 \end{eqnarray} 
 In this case we get for large $z$, $f_{\ell m}\sim z^{-\ell-1}$ since $n=0$ is the dominant term. It is easy to see that 
 \begin{eqnarray}
 q_{\ell m}\sim \frac{r^{\ell+1}}{t^{2\ell+2}}, 
 \end{eqnarray}
 so that 
\begin{eqnarray}
 Q_{\ell m}\sim \frac{1}{t^{2\ell+3}},
 \end{eqnarray} 
 and  \cite{Price:1971fb}
 \begin{eqnarray}
 h_{\ell m}^{(1)}\sim \frac{1}{t^{2\ell+3}}, \qquad 
 h_{\ell m}^{(0)}\sim \frac{1}{t^{2\ell+2}}.
 \end{eqnarray}

 \subsection{Quantum mechanical analogy}
\noindent
Let us note that the we can write Eq. (\ref{flm}) as
\begin{eqnarray}
H_\ell f_{\ell m}=0,
\end{eqnarray}
 where 
\begin{eqnarray}
H_\ell=(1-z^2)^2 \frac{{\rm d}^2}{{\rm d} z^2} 
-2z (1-z^2) \frac{{\rm d}}{{\rm d} z} +\ell(\ell+1)(1-z^2)-a^2. 
\end{eqnarray}
If we define  operators $L_\ell^\pm$  as
\begin{align}
L_\ell^+=& (z^2-1) \frac{{\rm d}}{{\rm d} z} +(\ell+1)\,z, \nonumber 
\\
L_\ell^-=& (z^2-1) \frac{{\rm d}}{{\rm d} z} -\ell \, z ,
\end{align}
then we can express $H_\ell $ as 
\begin{eqnarray}
H_\ell=L^+_{\ell-1}L^-_\ell+\ell^2-a^2=L^-_{\ell+1}L^+_\ell+(\ell+1)^2-a^2.
\label{Hq}
\end{eqnarray}
The operators $L_\ell ^\pm$ satisfy the relations 
\begin{eqnarray}
H_{\ell+1}L^+_\ell=L^+_\ell H_\ell, \qquad  H_{\ell-1}L^-_\ell=L^-_\ell H_\ell,
\end{eqnarray}
and therefore, $L_\ell^\pm$ can be considered as raising and lowering ladder operators. It is easy to see that we can define conserved quantities $C_\ell$ 
\begin{eqnarray}
C_\ell=L_{\ell-1}^+C_{\ell-1} L_\ell^-
\end{eqnarray}
such that 
\begin{eqnarray}
[C_\ell,H_\ell]=0. \label{CH}
\end{eqnarray}
The first quantity $C_0$ in this hierarchy is defined as 
\begin{eqnarray}
C_0=L^-_0=(z^2-1) \frac{{\rm d}}{{\rm d} z}, 
\end{eqnarray}
which satisfies 
\begin{eqnarray}
[C_0,H_0]=0, \qquad H_0=(1-z^2)^2 \frac{{\rm d}^2}{{\rm d} z^2} 
-2z (1-z^2) \frac{{\rm d}}{{\rm d} z} +a^2.
\end{eqnarray}
As we have already mentioned above,   the solution of the static problem of Eq. (\ref{qq}) is $r^{\ell+1}$, so that $H_\ell $ should have a solution, which for large $z$ scales like $z^{-\ell-1}$. Since the 
solution of 
\begin{eqnarray}
    L^+_\ell \phi=0, 
\end{eqnarray}
is solved by 
\begin{eqnarray}
    \phi= (z^2-1)^{-(\ell+1)/2},
\end{eqnarray}
which has the correct 
$z^{-\ell-1}$ at large $z$;  inspection of Eq. (\ref{Hq}), reveals that 
\begin{eqnarray}
    a=\ell+1,
\end{eqnarray}
as we found above, 
in order $H_\ell$ to have the correct scaling solution $z^{-\ell-1}$ at large $z$.

\section{Nonlinear tails: The retarded Green function}\label{supp:2}
\noindent
Let us consider the differential equation for a generic scalar $\Psi(r,t)$
\begin{eqnarray}
\frac{\partial^2\Psi(t,r)}{\partial t^2}-\frac{\partial^2\Psi(t,r)}{\partial r^2}(t,r)+\frac{\ell(\ell+1)}{r^2}\Psi(t,r)=0,
\end{eqnarray}
where for large $r$ we confuse $r_*$ with $r$.
The corresponding Green function $G_F(t,r;t',r')$ satisfies
\begin{eqnarray}
\frac{\partial^2G_F}{\partial t'^2}-\frac{\partial^2G_F}{\partial r'^2}
+\frac{\ell(\ell+1)}{r'^2}G_F=\delta(t-t') \delta(r-r'). 
\end{eqnarray}
Expanding $G_F(t,r;t',r')$  in Fourier modes as 
\begin{eqnarray}
G_F(t,r;t',r')=\frac{1}{\sqrt{2\pi}}\int_{-\infty}^\infty {\rm d} \omega e^{i \omega (t-t')}\widetilde{G}_F(\omega,r,r'), \label{gf}
\end{eqnarray}
we find that Green function in the frequency domain $\widetilde{G}(\omega,r,r')$ satisfies 
\begin{eqnarray}
\widetilde{G}_F''+\left(\omega^2-\frac{\ell(\ell+1)}{r^2}\right)\widetilde G_F=-\delta(r-r'), 
\end{eqnarray}
where the superscript ${}^{'}$ denotes derivative with respect to $r$. The homogeneous problem satisfies
\begin{eqnarray}
\phi''+\left(\omega^2-\frac{\ell(\ell+1)}{r^2}\right)\phi=0,
\label{PP}
\end{eqnarray} 
which is just the Schr\"odinger equation for the scattering of stationary states in an $1/r^2$, corresponding to 1D conformal quantum mechanics. 
Eq. (\ref{PP}) is solved by the modes 
\begin{eqnarray}
\phi_{1}(\omega,r)=\sqrt{\frac{2\omega}{\pi}}j_{\ell}(\omega r), \qquad \phi_{2}(\omega,r)=\sqrt{\frac{2\omega}{\pi}}n_{\ell}(\omega r), 
\end{eqnarray}
where $j_\ell$ and $n_\ell$ are spherical Bessel functions. Then, the Green function $\widetilde{G}(\omega,r,r')$ turns out to be 
\begin{eqnarray}
\widetilde{G}_F(\omega,r,r')=\omega r r' \bigg( j_\ell(\omega r) n_\ell(\omega r') \, \theta(r-r')+ j_\ell(\omega r') n_\ell(\omega r) \theta(r'-r)\,  \bigg),
\end{eqnarray}
so that the Green function in the time domain for $r>r'$ is given by 
\begin{eqnarray}
G_F(t,r;t',r')=\frac{1}{\sqrt{2\pi}}\int_{-\infty}^\infty {\rm d} \omega e^{i \omega (t-t')}\omega r r'  j_\ell(\omega r) n_\ell(\omega r'), \qquad r>r'.
\end{eqnarray}
By using the following representation of the spherical Bessel functions 
\begin{eqnarray}
j_\ell(x)=(-x)^\ell\left(\frac{1}{x}\frac{{\rm d}}{{\rm d} x}\right)^\ell \frac{\sin x}{x}, \qquad n_\ell(x)=-(-x)^\ell\left(\frac{1}{x}\frac{{\rm d}}{{\rm d} x}\right)^\ell \frac{\cos x}{x}, 
\end{eqnarray}
we find that 
\begin{eqnarray}
G_F(t,r;t',r')=r^{\ell+1}{r'}^{\ell+1} \left(\frac{1}{r}\frac{{\rm d}}{{\rm d} r}\right)^\ell  \left(\frac{1}{r'}\frac{{\rm d}}{{\rm d} r'}\right)^\ell    \frac{1}{r r'}\frac{1}{\sqrt{2\pi}}\int_{-\infty}^\infty {\rm d} \omega e^{i \omega (t-t')} \frac{\cos(\omega r)\sin(\omega r')}{\omega^{2\ell+1}}.  \label{GG}
\end{eqnarray}
The integral in Eq. (\ref{GG}) can be performed in the complex $\omega$-plane using the fact that it posseses a pole of order $(2\ell+1)$ at $\omega=0$. The result for $0<|r-r'|<t-t'<r+r'$ is 

\begin{eqnarray}
I(t,t',r,r')&=&\frac{1}{\sqrt{2\pi}}\int_{-\infty}^\infty {\rm d} \omega \,e^{i \omega (t-t')} \,\frac{\cos(\omega r)\sin(\omega r')}{\omega^{2\ell+1}}\nonumber\\
&=&\sqrt{\frac{\pi}{2}} \frac{(-1)^\ell}{4\cdot 2\ell!}\bigg[-((t-t')+r-r')^{2\ell}+((t-t')-r+r')^{2\ell}+((t-t')-r-r')^{2\ell}+((t-t')+r+r')^{2\ell}\bigg]. \nonumber\\
&&
\end{eqnarray}
Therefore, we have 
\begin{eqnarray}
G_F(t,r;t',r')=r^{\ell+1}{r'}^{\ell+1} \left(\frac{1}{r}\frac{{\rm d}}{{\rm d} r}\right)^\ell  \left(\frac{1}{r'}\frac{{\rm d}}{{\rm d} r'}\right)^\ell    \frac{I(t,t',r,r')}{r r'}. \label{GG1}
\end{eqnarray}
One can further proceed and evaluate the derivatives above,  but the resulting expressions are  not particularly illuminating. Instead, (\ref{GG1}) can be used to find easily the Green function  for any $\ell$. For example, for $\ell=2$ we get 
\begin{eqnarray}
G_F(t,r;t',r')=\sqrt{\frac{\pi}{2}} \frac{3(t-t')^4 +3 r^4+2 r^2 {r'}^2+3 {r'}^4-6(t-t')^2(r^2+{r'}^2)}{16 r^2 {r'}^2} 
\end{eqnarray}
which agrees with the result in Ref. \cite{Ling:2025wfv}.
For $\ell=4$, we get

\begin{eqnarray}
G_F(t,r;t',r')&=&\sqrt{\frac{\pi}{2}}\frac{1}{256\, r^4 {r'}^4}\left[35 r^8+20 r^6(r'^2-7(t-t')^2)+20 r^2(r'^2-7(t-t')^2)(r'^2-(t-t')^2)+35(r'^2-(t-t')^2)^4\right.\nonumber\\
&+&\left. 6r^4(3 r'^4-30r'^2(t-t')^2+35(t-t')^4)\right]. 
\end{eqnarray}

\end{document}